\documentclass[twocolumn,superscriptaddress,showpacs]{revtex4}
\usepackage{graphicx}
\begin{document}

\title{Effective non-additive pair potential for lock-and-key interacting particles: the role of the limited valence.}

\author{Julio Largo}\affiliation{ {Dipartimento di Fisica and
  INFM-CNR-SOFT, Universit\`a di Roma {\em La Sapienza}, Piazzale A. Moro  2, 00185 Roma, Italy} }
  % \affiliation{ {Dipartimento di Fisica,  Universit\`a di Roma {\em La Sapienza}, Piazzale A. Moro 2, 00185 Roma, Italy} }
\author{Piero Tartaglia} \affiliation{ {Dipartimento di Fisica and
  INFM-CNR-SMC, Universit\`a di Roma {\em La Sapienza}, Piazzale A. Moro  2, 00185 Roma, Italy} }
\author{Francesco Sciortino} \affiliation{ {Dipartimento di Fisica and
  INFM-CNR-SOFT, Universit\`a di Roma {\em La Sapienza}, Piazzale A. Moro  2, 00185 Roma, Italy} }

\begin{abstract}
Theoretical studies of self-assembly processes and condensed phases in colloidal systems are often based on effective inter-particle potentials. Here we show that developing an effective potential for particles interacting with a limited number of  ``lock-and-key'' selective bonds (due to the specificity of  bio-molecular interactions) requires --- beside  the non-sphericity of the potential  ---  a (many body) constraint that prevent multiple bonding on the same site.   We show the importance of retaining both valence and bond-selectivity by developing, as a case study, a simple effective potential describing the interaction between colloidal particles coated by  four single-strand DNA chains.
\end{abstract}
%\pacs{82.70.Dd, 82.70.Gg, 61.20.Ja: Version: Saturday Evening  }
%82.70.Dd (colloids) , and 82.70.Gg (gels) 61.20.Ja Computer simulation of liquid structure

\maketitle

Description of multi-scale phenomena in self-assembling systems and understanding of their collective behavior requires a coarse graining process in which the irrelevant degrees of freedom are integrated out in favor of an effective re-normalized interaction potential~\cite{likos-review}. Colloidal systems are  often described by this type of approach, in which the solvent and its chemical-physical properties are encoded into parameters of the resulting effective inter-colloid particle potential~\cite{russel-book}.   
%The DLVO~\cite{dlvo} potential between charged colloids or the Asakura-Osawa~\cite{ao,Vrij} potential describing the attraction between colloidal particles induced by depletion interactions are two of the most famous examples.
  Thermodynamic (and often also dynamic)  bulk behavior of several complex particles, including polymers, star-polymers, dendrimers, micelles, 
telecheric polymers~\cite{louis2000,likosstar,micelle,2006watermelon,pierleoni,zaccamicelle} has been interpreted on the basis
of effective pair-wise potentials,  function only of the  distance between particle centers. 
%More recently, in describing   the self-assembly properties of block-copolymer  in micelles and, under specific conditions, in super-micelle structures, the procedure for determining the effective potential has been iterated twice, focusing progressively on the self-assembly of the polymers in micelles and, with a further integration step, from independent to structured  micelles. 
%In most cases, the resulting effective pair-wise potential  is a function of the  distance between particles centers (i.e. angular information are integrated out). 

 Very recently, design of new materials has started to capitalize on the
possibilities offered by complementary ``lock-and-key'' selective interactions~\cite{niemeyer},  aiming to achieve a much better control on the resulting bonding pattern and on the  mechanical properties of the self-assembled bulk material. In this Letter we show that  effective potentials for these new particles must  account for both the valence (maximum number of bonds) and for the selectivity of the bonding process.  While at the microscopic level specificity in the bonding interactions is pair-wise additive --- it  builds on  steric incompatibilities~\cite{Werth1}, e.g. on excluded volume effects which  prevent the possibility of multiple interactions at the same bonding site --- we find that the  conservation of this important feature at the level of the  coarse-grained effective potential may require the break-down of the pair-wise additive approximation.

\begin{figure}[h]
\begin{center}
 \includegraphics[height=6cm, clip=true]{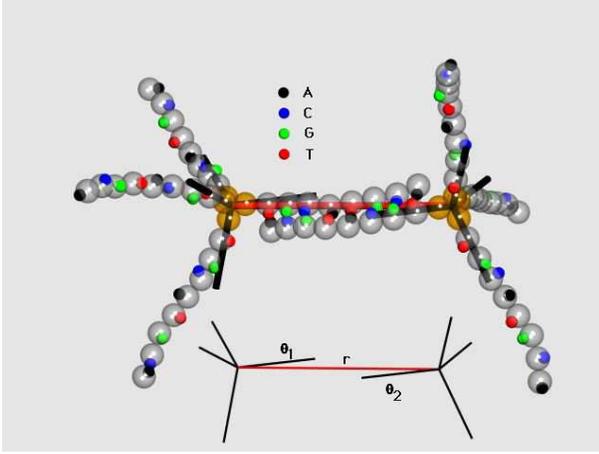}
\caption{ Drawing of two macromolecules in a bonded configuration.  Each  macromolecule is composed by a tetrahedral hub (four tetrahedrally-oriented orange spheres)  to which four single strands of eight {\it bases} each (grey spheres) are anchored.   Each base is decorated by a short-range attractive site (small spheres) encoding the base type (A,T,G,C).  Base complementary is enforced by selective attractive interactions between A and T and between G and C sites. In the process of developing a coarse-grained model, we associate a rigid tetrahedron (oriented according to the central core of the particle) to each macromolecule and  focus on the center-to-center distance $r$ and on the two angles $\theta_1$ and $\theta_2$, defined in the figure. 
In this Letter we use as unit of length the diameter of the larger spheres  (orange or grey) and as a unit of energy the  A-T (or the equivalent G-C) binding energy. Number density is defined as the total number of  large spheres (36 for each macromolecule) divided by the volume.}
\label{fig:figura}
\end{center}
\end{figure}

As a working example, we discuss the development of an effective potential for 
a macromolecule composed by a  core particle decorated by a small number of short  single strands (ss) DNA\cite{STEWART_04} to show (i)  the importance of properly accounting for valence and bond selectivity;  (ii) the need for a non-pair-wise additive effective potential. The full model is graphically explained in Fig.~\ref{fig:figura} (and the chosen length and energy units), and more in detail Ref.~\cite{jpcmstarr,langmuir}.   
A semi-rigid chain of eight particles, modeling a  8-base ss-DNA, is attached 
to each of the four particles composing a central tetrahedral core. 
Each of the 32 base particles carries an additional  interaction site 
(labeled A,T,G or C)  which can bind only with complementary site types
(only A-T or G-C binding are allowed), resulting in a total of  68 interaction sites per macromolecule.  The chosen  base sequence is palindromic, so that  all eight bases can be simultaneously  paired  when ss-arms of different macromolecules approach in the correct orientation.  All site-site interactions are pair-wise additive.  
The phase diagram of this model   shows a region of gas-liquid phase separation, but only at small number densities $\rho$. The liquid region is characterized by the presence of a four-coordinated network, whose dynamics progressively slows down on cooling, parallel to the formation of an open gel-like structure~\cite{langmuir}.

\begin{figure}[t]
\begin{center}
\includegraphics[height=10cm, clip=true]{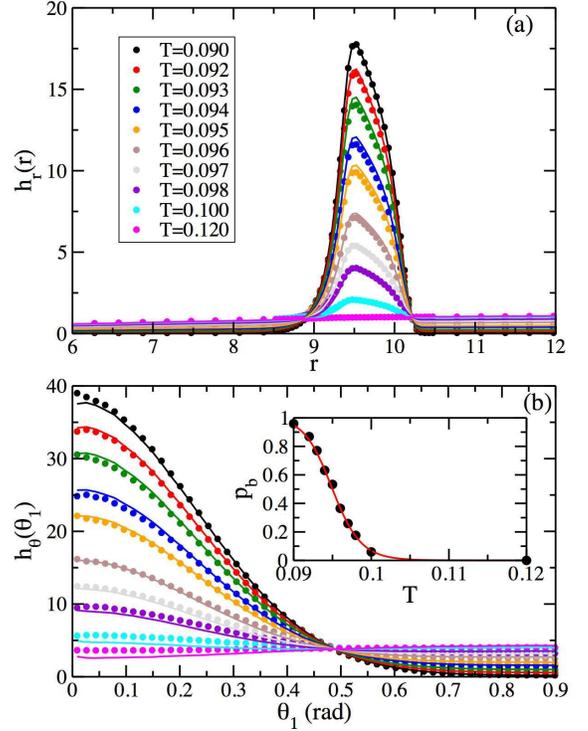}
\caption{Results from the simulation of two macromolecules in a box of side $L=25$:
center-to-center probability $h_r(r)$ (a) and angle $\theta_1$ distribution $h_{\theta}(\theta_1)$   ($h_{\theta}(\theta_2)=h_{\theta}(\theta_1)$ for symmetry).  Symbols are numerical results for different $T$. Lines are the corresponding
expressions provided by Eq.~\ref{eq:prt} after integration over $\theta_1$ and $\theta_2$ in panel (a) or integration over $r$ and $\theta_2$ in panel (b). The inset shows the $T$ dependence of $p_b$ and its fit to a two-state model (see text)
with $\Delta E=4.94$ and $\Delta S/k_B=52$ ($k_B$ is the Boltzmann constant).}
\label{fig:gr2comp}
\end{center}
\end{figure}

To evaluate the effective potential we simulate, using  Monte Carlo methods, a system composed of  two macromolecules in a  box 
%of side $L=25$  
for several temperatures $T$, covering the $T$-region where ss-DNA pairs. 
From a large ensemble of equilibrium configurations we extract the properties of the degrees of freedom to be retained 
in the coarse-grained model: (i)  the normalized probability $h_r$ of finding the two particles at relative center-to-center distance $r$ (i.e. in such a way that $\frac{1}{L^3} \int 4 \pi h_r r^2 dr = 1$) 
(ii) the normalized probabilities  $h_{\theta_1}$ and $h_{\theta_2}$  of the angle $\theta_1$  and $\theta_2$ between $\vec r$ and the closest ss-DNA arm on each particle (so that $\frac{1}{2} \int_0^\pi  h_\theta sin \theta d\theta = 1$). The variables  $r$, $\theta_1$ and $\theta_2$ are defined  in Fig.~\ref{fig:figura}.
The $T$-dependence of $h_r(r)$ and $h_\theta(\theta_1)$  is shown in Fig.~\ref{fig:gr2comp}. On cooling, angular and spatial correlations build up, reflecting the formation of a stable double-strand configuration, with a characteristic center-to-center distance of the order of the DNA arm length ($8.5 \lesssim r \lesssim 10.5 $)  and a preferential  orientation for linear bonding ($\theta_1 \approx 0 $). 
At the highest  and lowest investigated $T$, $h_r$ and $h_\theta$  describe, to a good approximation, the fully  non-bonded    ($nb$) and fully bonded  ($b$) state, respectively.   The  non-bonded  high $T$  state is characterized by essentially unstructured  $h_\theta^{nb}$ and $h_r^{nb}$, except for the $r$-region close to the origin where excluded volume interactions play a role.
The function $h_\theta^{b}$  is well described by a gaussian distribution 
$h_\theta^{b}=\frac{A}{\sqrt{2\pi \sigma^2}} e^{-\frac{\theta^2}{2 \sigma^2}}$  
(where the normalization constant  is $A= 22.32$ and the best-fit  $\sigma^2=0.052$ rad$^2$). A similar  angular distribution has been recently chosen to  model self-assembly of patchy particle\cite{doye}. The distribution $h_r^{b}$ is essentially localized only between $8.5 \lesssim r \lesssim 10.5$, corresponding to the
bonded distance. We have not been able to provide a simple analytic best-fit expression
for $h_r^b$ and $h_r^{nb}$ and,  in the following, we have used their numerical values.
The sharp separation in configuration space between the bonded and the  non-bonded  states (due to the localized nature of the interaction and to its specificity) suggests to describe the distributions at intermediate $T$ as a linear combination of the $b$ and $nb$ distributions, weighted respectively by  the probability of being  bonded ($p_b(T)$) or non-bonded ($1-p_b(T)$).  Lines in Fig.~\ref{fig:gr2comp} show  that both $h_r$ and $h_\theta$ can indeed be described as linear combinations (with the same $p_b(T)$ coefficient) of the $b$ and $nb$ distributions.  Moreover, as shown in the inset of Fig.~\ref{fig:gr2comp}, the resulting $T$-dependence of  $p_b(T)$ is very well fitted by the two-state expected theoretical expression  $p_b(T)=1/(1+e^{-\frac{\Delta U - T \Delta S}{k_B T}})$, where $\Delta U$ and $\Delta S$ measure the change in energy and entropy associated to the formation of a double strand.
The important result is that the $T$-dependence decouples from the $r$ and $\theta$ dependence and it becomes  possible to write the probability of finding the two macromolecules at relative distance $r$ and angles $\theta_1$ and $\theta_2$
 as
\begin{eqnarray}
P(r,\theta_1,\theta_2,T)= \\ \nonumber 
p_b(T)\,P^{b}(r,\theta_1,\theta_2)+[1-p_b(T)]\,P^{nb}(r,\theta_1,\theta_2).
\label{eq:prt}
\end{eqnarray}
We approximate  $P^{b}(r,\theta_1,\theta_2)=h_r^{b}(r) h_{\theta}^{b}(\theta_1) h_\theta^{b}(\theta_2)$  and  analogously $P^{nb}$  for the non-bonded case. 
In this approximation, correlation between distances and angles within the
$b$ (or the $nb$) state  are neglected, since both $P^b$  and $P^{nb}$ are written as a product of their 
arguments.  We have tested that this is indeed a
good approximation. Correlation between angles and distances 
arising from the differences between $b$ and $nb$ states  (small $\theta_1$ and $\theta_2$ are found  when $r$ is within the bonding distance) are retained in $P$, which indeed can not be factorized. \\
 \begin{figure}[h]
\begin{center}
 \includegraphics[height=6.cm, clip=true]{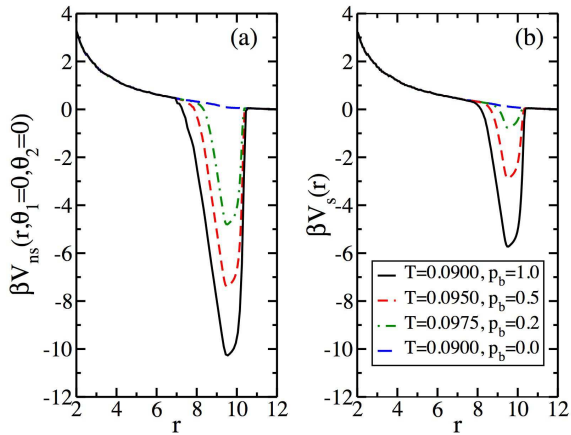}
 \caption{(a) Shape of the non-spherical effective potential $\beta V_{ns} (r)$ for given angles $\theta_1=0$ and $\theta_2=0$, and (b) the spherically averaged effective potential $\beta V_{s}(r)$, for different temperatures. }
\label{fig:potential}
\end{center}
\end{figure}

 Thus the four functions $h_r^{b}$, $h_{\theta}^{b}$, $h_r^{nb}$, $h_{\theta}^{nb}$ and  $p_b(T)$ fully specify  $P(r,\theta_1,\theta_2,T)$ and make it possible to 
model the macromolecule as a particle decorated by four tetrahedrally oriented arms and  evaluate  an effective potential  between these particles.  The simplest approach is to neglect the angular correlation and develop a spherical ($s)$ effective potential $V_{s}(r,T)$, integrating out the angular degrees of freedom as
\begin{equation}
\beta V_{s}(r,T)=- \ln{ \frac{  p_b(T)h_r^{b}(r)+[1-p_b(T)] h_r^{nb}(r) }
                              {p_b(T)h_r^{b}(r_\infty)+[1-p_b(T)] h_r^{nb}(r_\infty)}}.
\end{equation}
Here $r_\infty$ is any distance larger than the range of interaction of the two molecules (so that $V_{s}(r,T)=0$ at large distances).  A more accurate  approach requires retaining information on the angular orientations,  defining a non-spherical ($ns$) angular dependent potential $V_{ns}$ as
\begin{equation}
\beta V_{ns}(r,\theta_1,\theta_2,T)=- \ln{ \frac{P(r,\theta_1,\theta_2,T)}{P(r_\infty,\theta_1,\theta_2,T)}.
}
\end{equation}
By construction, both $V_s$ and $V_{ns}$ accurately describe the radial distribution function 
(and, in the case of $V_{ns}$, also the angular distribution function) of the system in the limit of
vanishing density at all $T$.  The radial shape of the two resulting potentials is shown in Fig.~\ref{fig:potential}, $V_{ns}$ has a deeper minimum than $V_s$, to compensate
for the reduction of the solid angle associated to bonding.

 \begin{figure}[h]
\begin{center}
 \includegraphics[height=6.cm, clip=true]{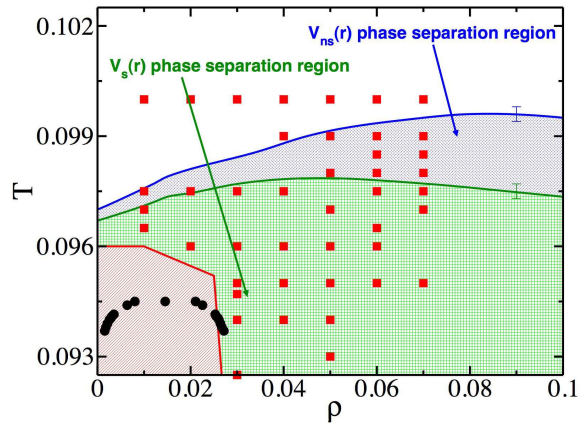}
 \caption{Phase diagram of the model for colloidal particles coated by  four single-strand DNA chains. 
   The dashed (red) area approximately indicates the region where phase-separation is expected for the model studied in Ref.~\cite{langmuir}  (estimates based  on  the extrapolated behavior of the low wave-vector  structure factor).  Filled squares, outside the dashed area, denote the stable state points studied in Ref.~\cite{langmuir}. The graph also shows the boundary between
   homogeneous and phase-separated states for $V_s$ and $V_{ns}$.  Note that most of the filled squares are located in a region of the phase-diagram where phase separation is observed in $V_s$ or $V_{ns}$. The filled circles indicate the gas-liquid  coexistence  for the valency constrained effective model.  }
\label{fig:phase}
\end{center}
\end{figure}

 To assess the ability of the two potentials in reproducing the bulk behavior of the original model and 
to estimate the importance of the directional interactions, we compare the phase diagram of the two potentials,
with the (previously studied) behavior of the original model~\cite{jpcmstarr,langmuir}.   
Both $V_s$ and $V_{ns}$ perform very badly,  predicting a very wide region of liquid-gas instability, which includes the region where the  tetrahedral network  of bonded particles is known to be the stable phase~(see Fig.~\ref{fig:phase}). 
The unstable gas-liquid region is even wider in the case of $V_{ns}$,
due to its  deeper minimum. The disagreement is so large that no state points can be chosen to compare the original and the effective models in the region where a network develops, since $V_s$ and $V_{ns}$ always generate a phase-separated configuration.\\
\begin{figure}[h]
\begin{center}
  \includegraphics[height=12cm, clip=true]{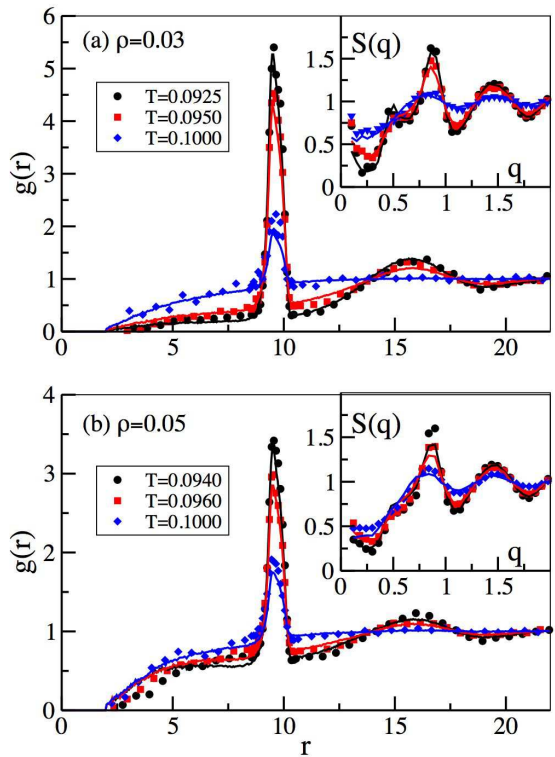}
 \caption{Radial distribution functions $g(r)$ for the original model (points, from Ref.~\cite{langmuir}) and the valency-constrained  effective potential  (lines) at
 $\rho=0.03$ (a) and $\rho=0.05$ (b). 
 Inset: structure factors $S(q)$ at the same state points. Note the increase in $S(q)$ at small $q$ when $\rho=0.03$,  signaling the proximity of the  limit of stability of the liquid phase. 
 }
\label{fig:grsq}
\end{center}
\end{figure}
The failure in reproducing the bulk behavior can be ascribed to the absence, in both  $V_s$ and $V_{ns}$,  of the lock-and-key character of the interaction. In the real model, once a ss-DNA is linked to a complementary strand of a different particles, it is not available for further bonding.  The specificity of the interaction implies that 
each ss-DNA can not be simultaneously bonded to more than one other arm. This important rule is not enforced 
 neither  by $V_s$ (which allows for a large number of bonded neighbors)  nor by $V_{ns}$ since the 
width of the angular part of $V_{ns}$ does not prevent multiple bonding at the same arm direction.  
We are forced to abandon the pair-wise
additivity approximation and develop a new potential
which enforces the bond selectivity and the valence (so that each particle
can not be engaged in more than four bonds).  To this aim, we 
 complement $V_{ns}$ with an ad-hoc rule imposing that 
each  arm  can be engaged only in
one bond, very much in the same spirit as the maximum valence models~\cite{Speedy}.  From a computational point of view, 
during the simulation, we retain a list of all pairs of bonded arms and make sure that each arm is involved at most in one bonded interaction.  
The resulting effective valency-constrained non-spherical model is able to generate homogenous equilibrium states in the expected region of the phase diagram. This opens the possibility  to compare  the predictions of the effective model with  those  of the original model~\cite{langmuir},  at the same
$T$ and $\rho$.  A comparison in real and in reciprocal space is 
shown in Fig.~\ref{fig:grsq}.  The limited-valence effective model
is able to  reproduce with satisfactory accuracy the structure of the tetrahedral network: both the peculiar  ratio between the position of the  first and second peak in the center-center radial distribution function $g(r)$ as well as the development of a pre-peak in the corresponding structure factor $S(q)$ on cooling are well captured by the effective potential. 
To locate precisely the gas-liquid coexistence for the limited-valence effective potential we perform umbrella sampling grand canonical Monte Carlo simulations~\cite{frenkelsmit}  (and histogram re-weighting techniques~\cite{Wilding_96}).  The resulting gas-liquid coexistence curve is shown in Fig.~\ref{fig:phase}.  
%The number density of the liquid at coexistence ($\rho \approx 0.028$), if scaled by measuring distances in units  of the particle center-to-center distance,  approaches  the value of  0.65 characteristic of the scaled number density of the closed-packed diamond crystal. 
As compared to the unconstrained $V_s$ and $V_{ns}$, the region of gas-liquid instability is now confined to small $\rho$, opening up a window of densities (above the liquid coexistence branch) in which equilibrium liquid states can be accessed. 

In summary, we have shown that in the process of developing effective potentials for particles with low valence and with selective interactions (as in functionalized colloids) it is mandatory to account for the specificity of the interaction which intrinsically limit the maximum number of bonds that can be formed. The present study shows that colloidal particles with specific interactions constitute another class of colloidal systems for which pair additivity approximation breaks down in the effective potential, in addition to the recently reported case of  charge-stabilized colloidal crystal~\cite{maret}.
   % In the case of DNA-functionalized particles this constraint arises from the fact that once a ss-DNA is attached to its complementary string, it is impossible to form another link with any incoming additional ss-DNA.  This feature  is essential in developing effective potentials for selective lock-and-key inter-paticle interactions.
The importance of retaining, beside the directionality, the valence of the original model  is clearly evidenced by the  fact that  valence strongly controls the width of the gas-liquid coexistence~\cite{zaccagel,bianchi}.

We thank C. Likos, F.~W. Starr  and  E. Zaccarelli for valuable discussions.
We acknowledge support from Miur PRIN and  MRTN-CT-2003-504712.

\bibliographystyle{apsrev}
\bibliography{./biblio.bib,./altre.bib}

\end{document}